\documentclass[12pt,dvips]{article}

\usepackage{amssymb} 
\usepackage{graphicx}
\usepackage{latexsym}

\newcommand{\be}{\begin{equation}}
\newcommand{\ee}{\end{equation}}
\newcommand{\bea}{\begin{eqnarray}}
\newcommand{\eea}{\end{eqnarray}}
\newcommand{\srm}[1]{{\textrm{\scriptsize #1}}}
\newcommand{\DCI}{{\ensuremath{D_\srm{CI}} }}
\newcommand{\DOV}{{\ensuremath{D_\srm{Ov}} }}
\newcommand{\Order}[1]{\ensuremath{\mathcal{O}(#1)}}
\newcommand{\trace}{\ensuremath{\textrm{tr}}}

\begin{document}
%%%%%%%%%%%%%%%%%%%%%%%%%%%%%%%%%%%%%%%%%%%%%%%%%%%%%%%%%%%%%%%%%%%%%%%%%%%

%%%%%%%%%%%%%%%%%%%%%%%%%%%%%%%%%%%%%%%%%%%%%%%%%%%%%%%%%%%%%%%%%%%%%%%%%%%
% TITLEPAGE
%%%%%%%%%%%%%%%%%%%%%%%%%%%%%%%%%%%%%%%%%%%%%%%%%%%%%%%%%%%%%%%%%%%%%%%%%%%
\thispagestyle{empty}

\date{\today}
\title{
Experiences with dynamical chirally improved fermions}
\author{C. B. Lang\thanks{e-mail:christian.lang@uni-graz.at}, 
Pushan Majumdar\thanks{e-mail:pushan.majumdar@uni-graz.at},
Wolfgang Ortner\thanks{e-mail:wolfgang.ortner@stud.uni-graz.at}\\
\\
{Institut f\"ur Physik, Theoretische Physik,} \\
{Karl-Franzens-Universit\"at, Graz, Austria}}
\maketitle
\begin{abstract}
We simulate Quantum Chromodynamics (QCD) in four Euclidean dimensions 
with two (degenerate mass) flavors of dynamical
quarks. The Dirac operator we use is the so-called chirally improved Dirac operator. We
discuss the algorithm used for the simulation as well as the checks and some results on
lattices up to size $8^4$ for fermion mass parameters down to 0.1. This is the first
attempt to introduce dynamical quarks with the chirally improved Dirac operator.
\end{abstract}


%%%%%%%%%%%%%%%%%%%%%%%%%%%%%%%%%%%%%%%%%%%%%%%%%%%%%%%%%%%%%%%%%%%%%%%%%%%
% MAIN PART OF PAPER
%%%%%%%%%%%%%%%%%%%%%%%%%%%%%%%%%%%%%%%%%%%%%%%%%%%%%%%%%%%%%%%%%%%%%%%%%%%

\newpage
\section{Introduction}

Spontaneous breaking of the chiral symmetry is one of the key issues in QCD. For a
lattice study of this phenomenon it is desirable to have a formalism which maintains chiral
symmetry as much as possible. Massless lattice Dirac operators obeying the
Ginsparg-Wilson condition (GWC) \cite{GiWi82},
\be\label{GWC}
D^\dagger+D=2\,D^\dagger\,R\,D \;,
\ee
(with a local operator $R$), provide the weakest form of violation of chiral
symmetry: it is violated locally and restored in the continuum limit. Also,
since the smallest quark masses are a few MeV we have to approach the chiral
limit in realistic lattice simulations eventually. 

Fermion zero modes are related to topological properties of QCD and may be
important for spontaneous  chiral symmetry breaking in the chiral limit. It is
therefore of particular importance for simulations with dynamical fermions to
be able to approach this limit. At the  moment only fermions defined through
Dirac operators obeying the GWC (GW-fermions) seem to be suited for an approach
to the chiral limit. An important property of such operators  (\ref{GWC}) for
$R=1/2$ is that their eigenvalue spectrum lies on a unit circle centered at 1
in the complex plane. 

It is extremely costly to include the full fermion dynamics with such fermions
as compared to the simpler Wilson or staggered fermions formulation, with
presently available algorithms. In fact, there have been very few attempts to
use  GW-fermions in dynamical simulations
\cite{BoHeEd99,FoKaSz04,CuFrEs04,CuKrFr04,DeSc04} and all of them have been
exploratory in nature.

These studies have been concerned with overlap fermions \cite{NaNe} which are
the only known exact  GW-fermions. This property, however, also gives rise to
additional problems due to the discontinuous development of the operator
spectrum even for a continuously changing gauge field. With the overlap
operator, zero modes appear or disappear instantaneously accompanying an
overall change of the  Dirac operator spectrum. 

Computationally more economic solutions are domain-wall fermions
\cite{Ka92FuSh95},  fixed-point fermions\cite{HaNi94} or chirally improved
fermions \cite{Ga01GaHiLa00}. Even though these actions either need an extra
dimension or they have considerably more terms, they are typically a factor of
\Order{10} more expensive than simpler actions but still \Order{10} less
expensive than overlap fermions.

\subsection{Chirally improved fermions}

Here we will work with the chirally improved fermions. Quenched simulations 
for this action have demonstrated good chiral properties allowing for
pion masses down to $\sim$ 250 MeV. The ground state hadron spectrum has been determined, e.g., in 
\cite{GaGoHa03a} and for excited hadrons in \cite{BuGaGl04a}.
 
The chirally improved massless Dirac operator may be written as a truncated series of terms
\bea\label{DCIaction}
S_\srm{CI}&=& \sum_x \sum_P \,\bar\psi(x) \,\DCI(x,x+P) \,\psi(x+P)\nonumber \;,\\
\DCI(x,x+P) &=&  \sum_{\alpha=1}^{16}\, \Gamma_\alpha\, c^\alpha_P \,U(x,x+P)    \;.
\eea
The sum in the action runs over path shapes $P$ connecting $x$ with $x+P$ while
the sum over $\alpha$ in \DCI runs over all elements of the Clifford algebra. 
$U$ denotes the ordered product of link variables along this path. 

The massive operator we define as
\be
\DCI(m)=m+\DCI\;,
\ee
where $m$ denotes the dimensionless quark mass \footnote{Actually, due to a trivial
renormalization the correct  mass value is $m/(1+m/2)$ but  for simplicity of notation we
always refer to $m$ here.}, i.e., it is the valence quark mass $m_\srm{val}$ in the
quenched simulation and agrees with the sea-quark mass $m_\srm{sea}$ in the dynamical
case. 

Plugging (\ref{DCIaction}) into the GWC and truncating the system (number of
coefficients and equations) one obtains a set of algebraic relations for the
coefficients $ c^\alpha_P$. The lattice symmetries, invariance under charge conjugation and parity
as well as  $\gamma_5$-hermiticity are respected but the series is truncated at path
length 4 and only a subset of 19 coefficients has been considered. These
coefficients depend implicitly on the gauge coupling and have to be re-determined at
different values. The leading gauge coupling dependence is -- similar to tadpole
improvement -- coded in two parameters $z_s$ and $z_v$ which multiply the gauge
links in the formal expansion. Usually it is sufficient to take $z_s=z_v$ and one
adjusts this value such that the spectrum of $\DCI$, which approximately follows 
the GW-circle, passes through zero. Coefficients for the quenched simulation \cite{GaGoHa03a}
can be found in \cite{GaLa03}.

This is an important technical point: The coefficients of  \DCI depend on both, the
gauge coupling $\beta$ and the sea-quark mass $m_\srm{sea}$ and have to be
determined by adjusting $z_s$ and solving the algebraic equations resulting from the
GWC as discussed in \cite{Ga01GaHiLa00}.

Our Dirac operator is always defined on one-step hypercubic (HYP) 
smeared \cite{HaKn01}  gauge configurations in order to
reduce ultraviolet (UV) fluctuations. In that sense the definition 
of our \DCI includes the smearing step.

\subsection{L\"uscher-Weisz action with tadpole improvement}

Previous experience in quenched calculations showed that using the
L\"uscher-Weisz action with coefficients from tadpole improved perturbation
theory leads to nicer chiral properties for this  Dirac operator
\cite{Ga01GaHiLa00}.  In particular the spectrum of the Dirac operator at small
eigenvalues deviates less from the circular shape. For that reason we also use
that gauge action here; it reads
\bea\label{LWaction}
S_\srm{LW}&=&
-\beta_1\sum_\srm{plaq}\frac{1}{3}\,\textrm{Re \,\trace}\;U_\srm{plaq} 
-\beta_2\sum_\srm{re}\frac{1}{3}\,\textrm{Re \,\trace}\;U_\srm{re}
\nonumber \\
&&
-\beta_3\sum_\srm{tb}\frac{1}{3}\,\textrm{Re \,\trace}\;U_\srm{tb}\;,
\eea
where $U_\srm{plaq}$ is the usual plaquette term,  $U_\srm{re}$ are Wilson
loops of rectangular $2 \times 1$ shape and  $U_\srm{tb}$ denote loops of
length 6 along edges of 3-cubes (``twisted bent'' or ``twisted chair''). The
coefficient $\beta_1$ is the independent gauge coupling and the other two 
coefficients $\beta_2$ and $\beta_3$ are determined from tadpole-improved
perturbation theory. They have to be calculated self-consistently
\cite{AlDiLe95} from
\be\label{LWparams}
u_0=\left (\frac{1}{3}\,\textrm{Re \,\trace}\langle U_\srm{plaq}\rangle\right)^{\frac{1}{4}}
\;,\quad
\alpha=-\frac{1}{3.06839}\,\log\left(u_0^4\right)\;,
\ee
through
\be
\beta_2=\frac{\beta_1}{20\, u_0^2}\,(1+0.4805\,\alpha)\;,\quad
\beta_3=\frac{\beta_1}{u_0^2}\,0.03325\,\alpha \;.
\ee
Again, this determination should be done for each pair of couplings $(\beta,
m_\srm{sea})$.

We discuss here results for lattices up to size $8^4$  and sea-quark masses
down to 0.1; our emphasis lies on the method, although we do discuss effects of
dynamical fermions on lattice spacing and propagators. Like other studies for
GW-fermions, our  study also has an exploratory character hopefully on the way
towards implementing the approach for large scale simulation.

\section{The updating algorithm}

We simulate QCD with 2 flavors of quarks with degenerate sea-quark
mass  $m_\srm{sea}$ using the chirally improved Dirac operator. The action thus has the
form
\be\label{completeaction}
S[\phi,U] =S_\srm{LW}+\phi^\dagger(\DCI^{-1}(m_\srm{sea}))^{\dag}(\DCI^{-1}(m_\srm{sea}))\phi\;,
\ee
where $\phi$ is the usual pseudo-fermion field \cite{WePe81}.

As mentioned, our Dirac operator includes HYP-smearing of the gauge configuration. This
smearing procedure involves the projection of a general complex matrix into SU(3). This
operation is not differentiable and the use of the exact Hybrid  Monte Carlo (HMC)
method \cite{HMCpapers} is therefore ruled out. 

The algorithm we implement can be thought of as a variation of standard HMC. To ensure
detailed balance in HMC  one introduces auxiliary momenta $p$ (conjugate to $U$) and
defines  a Hamiltonian ${\cal H}$ by 
\be
{\cal H}=\frac{1}{2}\,\trace\, p^2 + S[\phi,U]\;,
\ee
where $S[\phi,U]$ is the original action of the theory. The molecular dynamics is 
driven by the Hamiltonian equations of motion and the pseudo-fermion field $\phi$ is
held fixed throughout the molecular dynamics trajectory. The final step is an accept/reject 
step with the acceptance probability $P_\srm{acc}$ given by
\be
P_\srm{acc}={\rm min}\left\{1,\frac{\exp(-{\cal H}_\srm{new})}{\exp(-{\cal H}_\srm{
old})}\right\}\;.
\ee 
The equilibrium distribution is determined  entirely by the action used in the
accept/reject step as long as the  molecular dynamics trajectories are reversible. The
molecular dynamics  evolution does not necessarily have to be generated by the same
action. Exploiting this  freedom we use a two-step algorithm in which  the first step
consists of making a proposal according to some simple (computationally cheap) action
development and the second step  is the accept/reject step with the original action. 

For the molecular dynamics step we define our Hamiltonian by 
\be
{\cal H}=\frac{1}{2}\,\trace\, p^2 + S_\srm{simple}\;,
\ee
where $S_\srm{simple}$ is a simpler, numerically cheaper action. 
Our acceptance probability nevertheless is given by   
\be
P_\srm{acc}={\rm min}\left\{1,\frac{\exp(-\{p^2+S[\phi,U]\}_\srm{new})}
{\exp(-\{p^2+S[\phi,U]\}_\srm{old})}\right\}
\ee
where $S$ now is the full, original action. Thus we have an exact algorithm and
need not worry about systematic biases. The central problem now is how to
generate proposals  for configurations efficiently, such that the path through
configuration space is as close as possible to that of the original action,
i.e., how to choose an efficient $S_\srm{simple}$. 

On our path to the finally chosen algorithm we tested various alternatives. An
obvious first choice for $S_\srm{simple}$ is the L\"uscher-Weisz gauge action
and the Wilson Dirac operator. The parameters of the Dirac operator  were chosen
to correspond approximately to the plaquette value and sea-quark mass
represented by our values chosen for \DCI in the accept/reject step.  We then also
tried to replace the Wilson Dirac operator by a truncated \DCI  including only
terms up to length 2. 

Finally we turned off the fermionic part of the molecular dynamics equations.  
At that point  our $S_\srm{simple}$ consisted only of the L\"uscher-Weisz gauge
action.  This turned out to be superior: it is faster and the inclusion of the
fermionic parts did not significantly improve the final acceptance.

\begin{figure}[t]
\begin{center}
\includegraphics*[width=8cm]{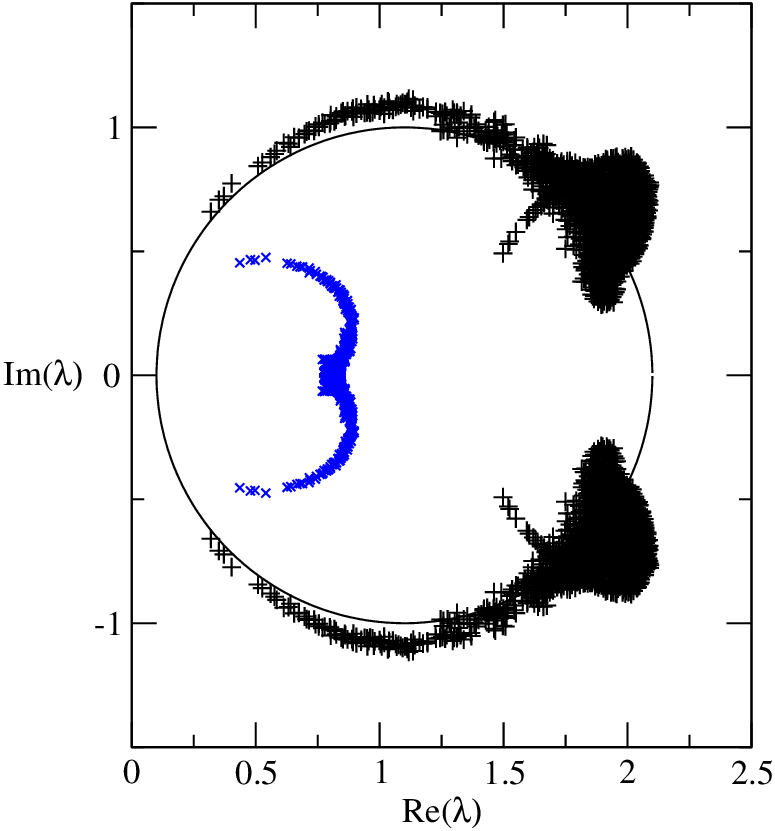}
\caption{We compare the eigenvalue spectrum of \DCI (+) with that of the reduced
operator $D_\srm{CI,r}$ ($\times$) for a typical configuration 
($4^4$ lattice, $\beta_1=7.4$, $m_\srm{sea}=0.1$); 
the latter are much closer to the value 1.}
\label{fig:spec_uv}
\end{center}
\end{figure}

The accept/reject step involves the ratio of determinants of the Dirac
operator. This is approximated by the stochastic estimator method inherent
in the pseudo-fermion formulation. It was pointed out
\cite{Ha99,AlHa02,ClKe04} that the noise in this stochastic estimation may
introduce artificial barriers on  the way through configuration space. 
Various methods to reduce the fluctuation of this estimate have been
discussed. Following these ideas  we introduced an  additional UV-filter
\cite{Fo99a}.  The basic concept is to reduce the spread of the eigenvalues
of the operator to be stochastically estimated. One defines a reduced
matrix
\be
D_\srm{r}=D \,\exp{(f(D))}\;,
\ee where $f(D)$ is chosen to be a polynomial in $D$ with coefficients 
such that the eigenvalues of $D_\srm{r}$ are concentrated around $z=1$ in
the complex plane.  In our acceptance step we need to compute 
\be
\frac{{\det}(D^{\dag}\,D)_\srm{new}}{{\det}(D^{\dag}\,D)_\srm{old}}=
\exp{\left(-2\,\trace\,f(D)_\srm{new}+2\,\trace\,f(D)_\srm{old}\right)}
\,\frac{{\det}(D_\srm{r}^{\dag}\,D_\srm{r})_\srm{new}}{{\det}(D_\srm{r}^{\dag}\,D_\srm{r})_\srm{old}}.
\ee
A complication for extended Dirac operators is to compute the
trace over polynomials of $D$. 
However for \DCI it is relatively straightforward to compute at least
$\trace(D)$ and $\trace(D^\dagger D)$ . We tried a polynomial
\be
f(D)=\gamma\,\trace(D)+\delta\, \trace(D^\dagger D)\;,
\ee
but found that introducing $\delta$ did not affect our results significantly. We
therefore made the simplest choice $f(D)=\gamma\,D$ with $\gamma=-0.477$. In Fig.
\ref{fig:spec_uv} we compare the eigenvalue spectrum of $\DCI$ with that of the
reduced operator  and find, indeed, that the reduced operator is closer to unity.

Our finally chosen updating method thus used this UV-filter combined with 
HYP-smearing and molecular dynamics equations using only the LW gauge  action.
Our trajectory lengths are 0.07 units of molecular dynamics time. In the actual 
implementation our trajectory consisted of a half step, a full step  and a final
half step in terms of the gauge field.

The acceptance itself is done in two steps. In the first step the value
$\exp\left(-2\,\gamma\,\trace\left(D_\srm{new}- D_\srm{old}\right)\right)$ is
calculated exactly along with the change in the kinetic energy and bosonic part
of the action. If this is accepted, the second -- more expensive -- step is the
stochastic estimation of the ratio of determinants of the reduced matrices. The
acceptance in the first step is $\sim 23\%$ and that of the second step is
$\sim 20\%$. Although this gives an overall acceptance rate of less than $5\%$,
one has to keep in mind that neither the first step nor the proposal involve
any inversion of the fermion matrix and both are therefore quite fast. The 
only time consuming step is the second. Thus the net efficiency is to be 
compared with an HMC close to $\sim 20\%$ acceptance rate, albeit with smaller 
trajectory length.

Since the inversion of the chirally improved operator is the most expensive
step in our algorithm we tried to increase its efficiency as much as possible.
It is well-known \cite{HMCpapers} that during the trajectory development in
standard HMC the inversion of the Dirac operator can be speeded up by using the
solution vector of the previous  step as the initial guess for the  current
step. Our case is slightly different. We do not invert the Dirac operator
during our trajectory development, but our trajectory lengths are smaller.
Therefore we expect that the new gauge field configuration $U_\srm{new}$ is
not too far away from the starting field configuration $U_\srm{old}$. Also we
note that the pseudo-fermion field $\phi$ is generated by
$\phi=\DCI(U_\srm{old})\,\xi$, where $\xi$ is a complex gaussian random vector.
Thus we have  $\xi=\DCI(U_\srm{old})^{-1}\,\phi$ and we use this vector $\xi$
as an initial guess for the inversion. Indeed this  choice reduces the
necessary number of matrix-vector multiplications by about 20\%.

Another speeding-up technique we use is to utilize the fact that we do not need to estimate the
determinant ratio exactly. 
If $\eta$ is the random number with which we want
to compare the ratio, then we need to check only if $(-\log\eta)$ is larger
than the change in action or not \cite{AlHa02}.   For the overlap operator this can be
implemented as follows.  The Metropolis accept/reject
step compares the  norm of solution vector $x=\DOV^{-1}\,\phi$ with
$(-\log\eta)$. At the $n$-th step of the bi-conjugate gradient routine, let the
solution vector be $x_n$ and the residual vector $r_n=\phi-\DOV\,x_n$. For
the overlap operator one knows that
\be
\frac{1}{2+m}|\phi|\leq|\DOV^{-1}\phi|\leq\frac{1}{m}|\phi|\;.
\ee
Then it is straightforward to show that 
\be
|x_n|^2-\frac{2}{m}|x_n||r_n|+\frac{1}{(2+m)^2}|r_n|^2\leq|x|^2\leq|x_n|^2+\frac{2}{m}|x_n||r_n|
+\frac{1}{m^2}|r_n|^2\;.
\ee
Since \DCI satisfies the GW-relation (with $R=1/2$) to a good approximation, we
assumed (and checked numerically) that its spectrum satisfies these bounds too.
So at every step we only need to compute the upper and lower bounds on $|x|^2$ to
see if $(-\log\eta)$ is inside that range or not. This reduces the number of
matrix vector multiplications typically by a factor 3. For a test of our
assumptions  we check the accuracy of this method by computing the  solution
vector to an accuracy of $10^{-12}$ randomly once in 20 updates on the average.
These tests never failed in our study.

We checked our numerics in several ways. Internal consistency of the program
was checked by verifying the reversibility of the molecular dynamics
trajectories. After a forward and  backward trajectory the final energy was
equal to the starting energy up to the precision  of our calculations (double
precision).  As another check of the possible problem of using a noisy
estimator  we compared the stochastic estimator for the ratio of determinant
with an exact evaluation (on the $4^4$ lattices). We found that the resulting
plaquette expectation value was equal within errors (less than 0.15\%)
whether we calculated the
determinant exactly or estimated it using the pseudo-fermions. 

Further checks, also on $4^4$ lattices, were to reproduce the quenched
plaquette values by turning off the  fermions, reproduce dynamical Wilson
results by replacing the chirally improved operator by the Wilson Dirac
operator in the accept/reject step and reducing the chirally improved  operator
to the Wilson Dirac operator by changing the coefficients.

We have discussed in the introduction that the \DCI parameters depend on the
normalization parameter $z_s$ which is adjusted  such that the massless
operator has eigenvalues running through zero. This parameter is a  function of
$\beta_1$ and $m$. Also the LW-action parameters $\beta_2$ and $\beta_3$ are
functions of  $\beta_1$ and $m$. All of them, $z_s(\beta_1,\, m)$,
$\beta_2(\beta_1,\,m)$ and $\beta_3(\beta_1,\,m)$ have to be determined
self-consistently by iterating the defining equations for  \DCI and the
LW-action.

To determine $\beta_2$ and $\beta_3$ self-consistently due to (\ref{LWparams})
we used a ``moving average'' of the plaquette, i.e., the average of the
plaquette  over a reasonably large interval of successive updates, and set it
to $u_0^4$.  The interval is shifted with new updates  by dropping the oldest
point and adding a new one.  The moving average has less fluctuations compared
to the original plaquette and in equilibrium  it is practically identical to
the plaquette average. Once equilibrium is reached we do not change the gauge
couplings any more. The final numbers are given in Table \ref{tab:tab0}.

\section{Results}

In the quenched case for the LW-action the lattice spacing values for various
values of $\beta_1$ have been determined in \cite{GaHoSc01,AlDiLe95a}. For
$\beta_1=7.6$ we have a lattice spacing $a\approx 0.19(1)$~fm, corresponding to
a lattice size of 1.5~fm  for the $8^4$ lattices in the quenched case.

Our final results are mainly from a simulation of the chirally improved Dirac
operator on a $8^4$ lattice with the tadpole improved L\"uscher-Weisz gauge
action at $\beta_1=7.6$ and quark mass parameter $m_\srm{sea}=0.1$. The
measured plaquette value for this run (cf. Table \ref{tab:tab0})  compares very
well  with the assumed plaquette value used for the determination of the
LW-action (\ref{LWaction}). Unless explicitly stated otherwise, all results 
refer to this run; it is a sequence of 120000 updates. With our average
acceptance rate of $\sim 5\%$ per update (i.e. per accept/reject step) and
individual trajectory lengths of 0.07 this corresponds to a total effective
molecular dynamics time of 420, counting only accepted steps. We allowed 40000
updates for equilibration and then saved a configuration every 2000 updates.
All our propagators and masses have been computed on 40 such configurations. 
In computing the propagators an additional complication, compared to the
quenched case, is that the masses of the sea-quarks and the valence-quarks
should agree; one therefore cannot use multi-mass solvers.

The total run time for equilibration and configuration generation was  about 2
weeks on a Linux cluster using 32 2.4 GHz Xeon CPU's.

\subsection{Equilibration}

\begin{figure}[t]
\begin{center}
\includegraphics*[width=8cm]{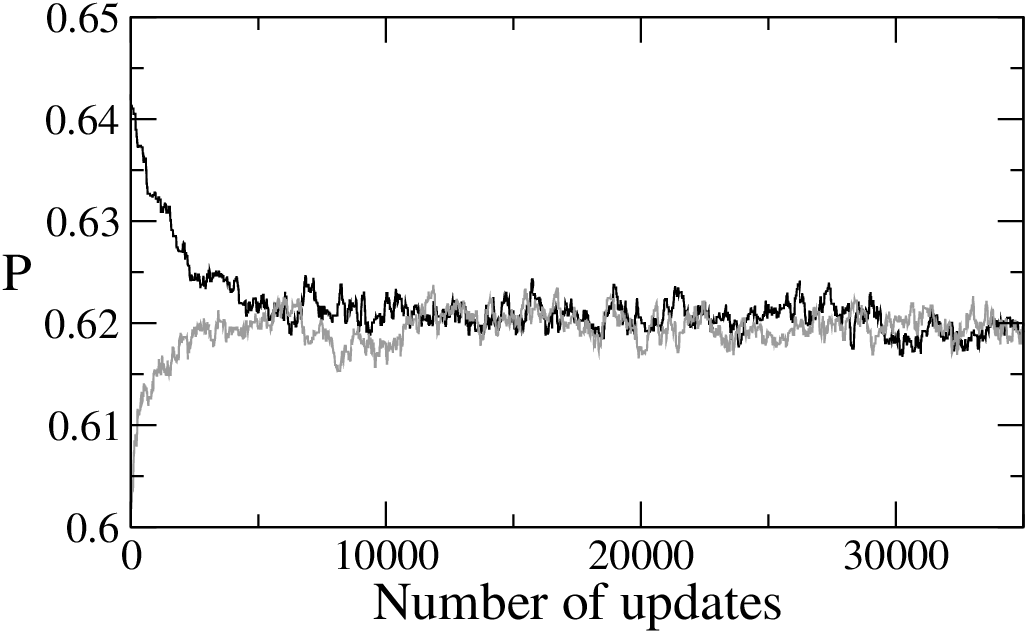}
\caption{Equilibration of the plaquette for lattice size $8^4$, 
$\beta_1=7.6$ and quark mass $m_\textrm{sea}=0.1$.}\label{fig:equil}
\end{center}
\end{figure}

Equilibration for the chirally improved operator is a rather slow process. Also
on a $4^4$ lattice we noticed that the number of matrix vector multiplications 
required were very high during the initial part of the cold start. This made
cold starts quite impracticable for larger lattices. On the $8^4$ lattice we
chose for our starting configurations quenched configurations  with  plaquette
values significantly higher and lower than the expected dynamical equilibrium 
value and let the two sequences converge. The plaquette history for such a
process is shown in Fig. \ref{fig:equil}, where we denote the plaquette value by 
$P=\frac{1}{3}\,\textrm{Re \,\trace}\langle U_\srm{plaq}\rangle$.

\begin{figure}[t]
\begin{center}
\includegraphics*[width=8cm]{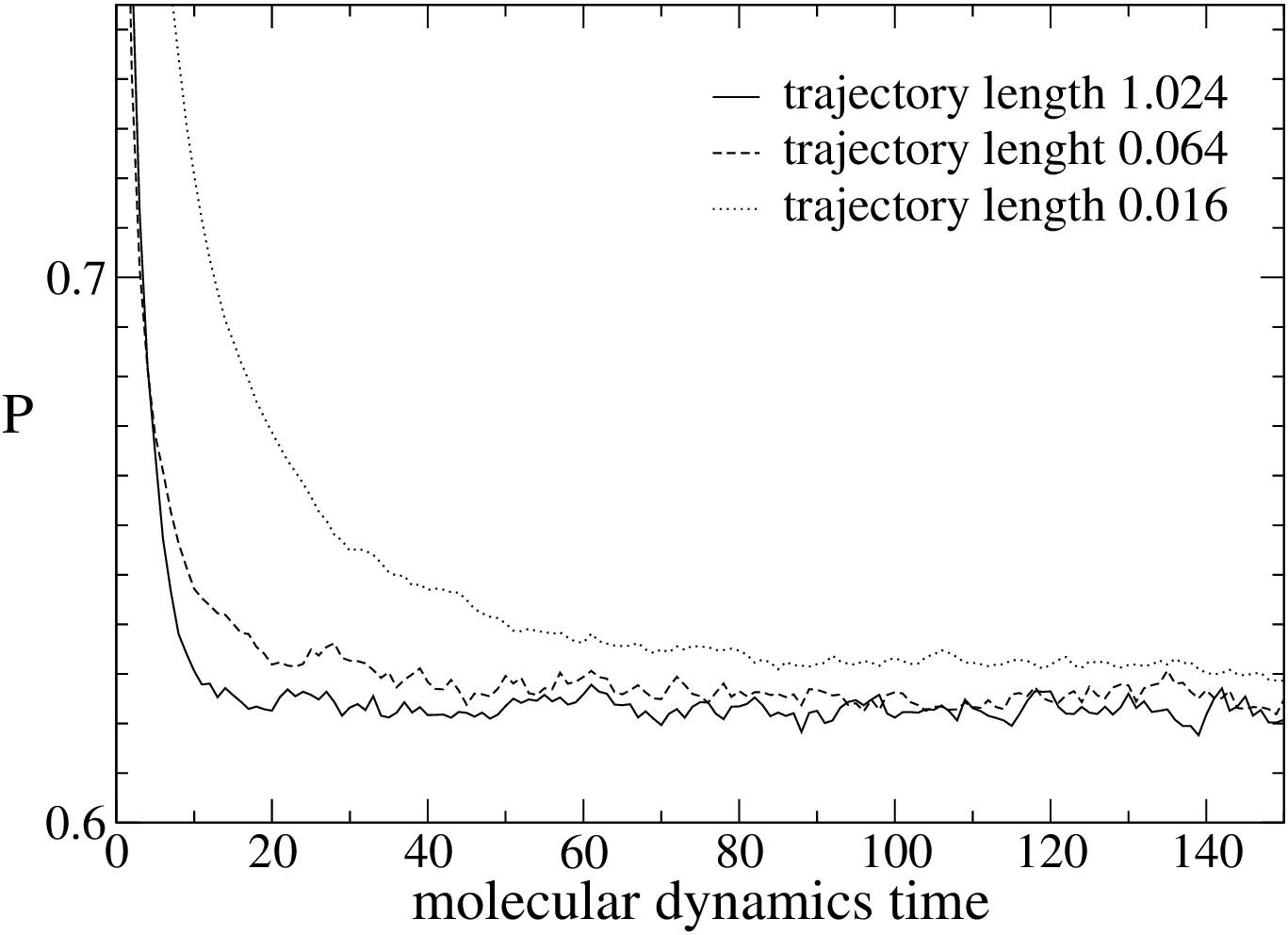}
\caption{Equilibration of the plaquette from cold start as discussed in 
the text. The three different 
curves correspond to trajectory lengths of 1.024, 0.064 and 0.016
in molecular dynamics time.}\label{fig:auto}
\end{center}
\end{figure}

Another important question is that of autocorrelation. Experience with exact
HMC shows that very small step sizes lead to rather large autocorrelation
times. We believe that with our trajectory length of $\Delta t=0.07$,  the
autocorrelation times are moderate. This is based on a quenched  study of the
autocorrelation time of the plaquette and we show the results in Fig.
\ref{fig:auto}.

From this figure we see that while the run with the trajectory length of 1.024 
falls fastest, the run with trajectory length 0.064 is not too different from
it after 20 time units whereas the run with length 0.016 is still quite far
away. Assuming that such a picture also holds for the dynamical case, we
conclude that our autocorrelation length is only moderately larger than for
standard HMC where one typically uses a trajectory length $\sim 1$. 

\begin{figure}[t]
\begin{center}
\includegraphics*[width=8cm]{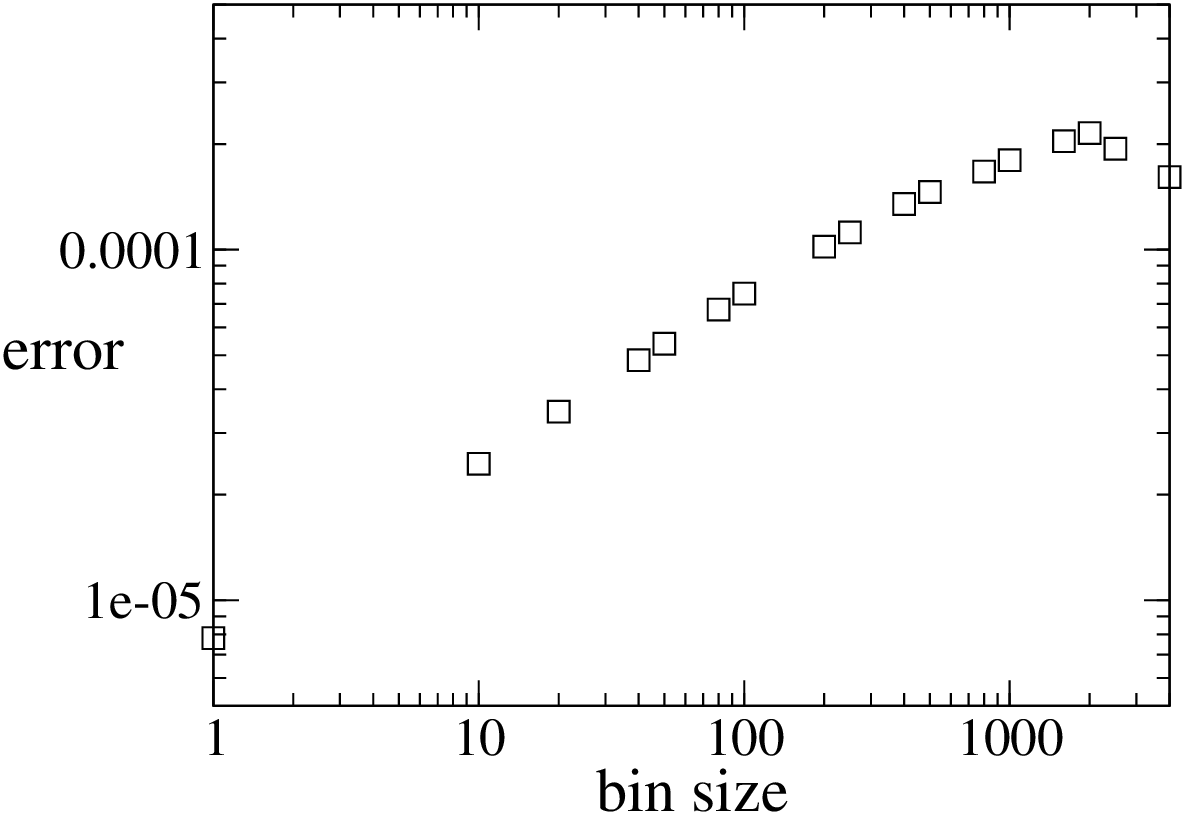}
\caption{Variation of the error for the plaquette with bin size
allowing for an estimate of the autocorrelation time ($8^4$ lattice,               
$\beta_1=7.6$ for the L\"uscher-Weisz gauge action.)}\label{fig:bin}
\end{center}
\end{figure}

To get an idea of the autocorrelation time in our runs, we did a binned error 
analysis for the plaquette. We plot the result in Fig. \ref{fig:bin}. As can be
seen  clearly from the figure, the maximal error is obtained around a bin size
of 2000, corresponding to an effective molecular dynamics time interval of 7.
We  take this to be an estimate of our autocorrelation time.
 
\begin{table}[ht]
\caption{Measured plaquette values for different masses; $u_0^4$ denotes the 
assumed plaquette value.}
\label{tab:tab0}
\begin{center}
\begin{tabular}{ccc|lllllll}
\hline  
$V$   & $\beta_1$ & $m_\srm{sea}$  & 0.1       & 0.5   & 2.0   & 10.0 & $\infty$ \\
\hline
$4^4$ & $ 7.4 $   & $U_\srm{plaq}$ & 0.608(1)  & 0.604(1) & 0.591(1)  &  0.579(1) & 0.556(2) \\
      &           & $u_0^4$        & 0.606     & 0.601    & 0.591     &  0.582    & 0.556 \\ 
\hline
$8^4$ & $ 7.6 $   & $U_\srm{plaq}$ & 0.6202(2) &       &       &      & 0.5824(1)\\
      &           & $u_0^4$        & 0.62      &       &       &      & 0.5825\\
\hline
\end{tabular}
\end{center}
\end{table}

On a $4^4$ lattice we also studied the dependence of the plaquette expectation
value on the quark mass. These studies were carried out for the L\"uscher-Weisz
action at $\beta_1=7.4$. The results are given in Table \ref{tab:tab0} together
with the values for the simulation on the $8^4$-lattice. Since switching on
dynamical fermions should in  leading order be equivalent to going to larger
$\beta_1$ in a quenched simulation, one expects that the plaquette value
increases for decreasing sea-quark mass. This is indeed what we observe.

\subsection{Spectrum of the Dirac operator}

\begin{figure}[t]
\begin{center}
\includegraphics*[width=8cm]{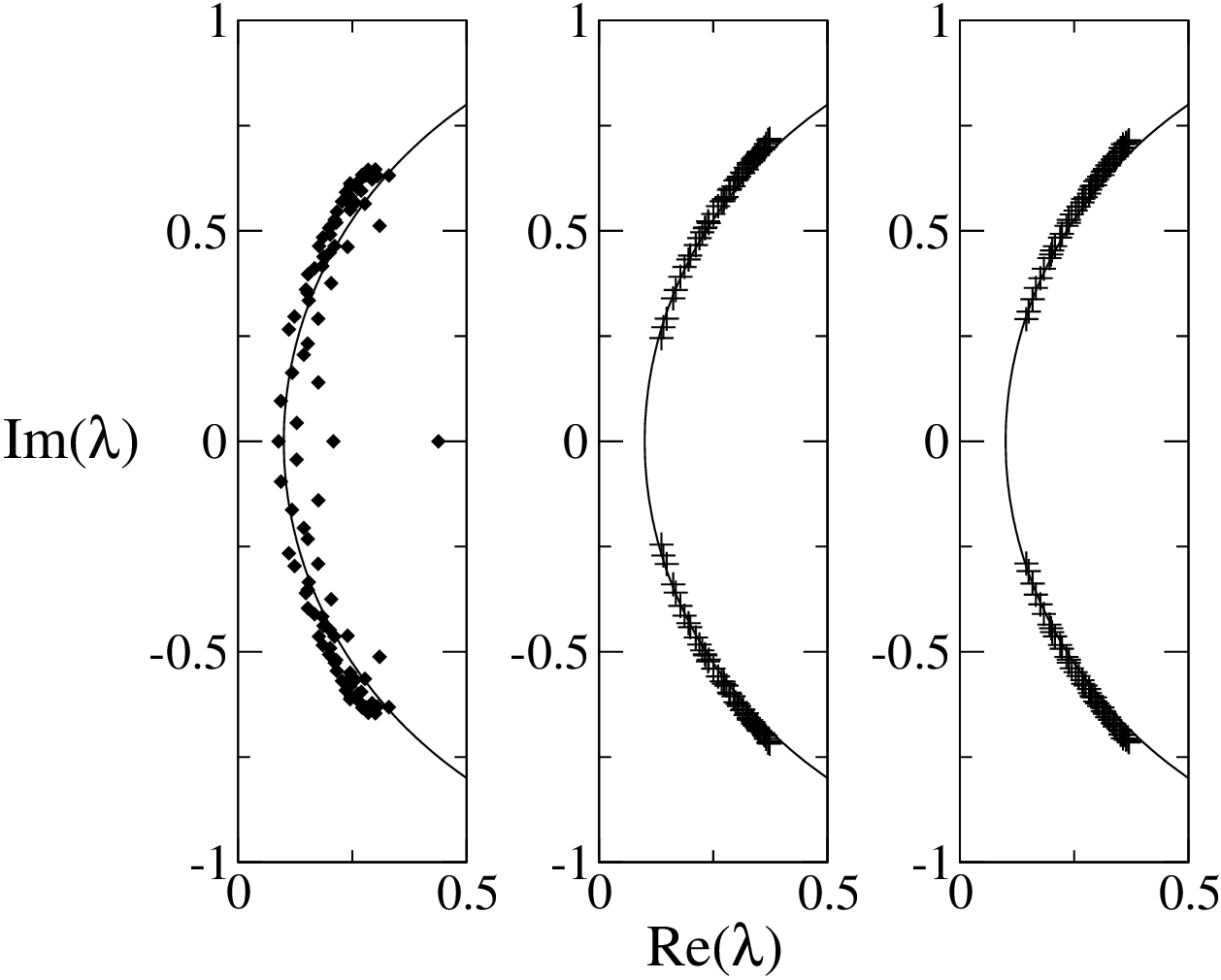}
\caption{Spectra of the chirally improved operator on a $8^4$ lattice. 
$\beta_1$ for the L\"uscher-Weisz gauge action was 7.6 and the quark mass 
$m=0.1$. These spectra are for three typical, randomly selected quenched (diamonds)
and dynamical (plus) configurations.
\label{fig:spec}}
\end{center}
\end{figure}

The chirally improved Dirac operator is an approximate solution of the 
Ginsparg-Wilson relation. In particular it has the property that the low lying 
spectrum is close to the Ginsparg-Wilson circle.  In order to verify that this
property was  preserved also with dynamical quarks, we plot the
first 100 eigenvalues for three equilibrium configurations in Fig.
\ref{fig:spec}, comparing the quenched with the dynamical situation. Compared
to the quenched case the spectrum in the dynamical case is closer to the
circle,  indicating smaller effective lattice spacing.

Zero modes of the Dirac operator are related to instantons. For the
chirally improved operator, in the quenched as well as the dynamical
case with large quark masses, zero modes are observed.  However for the mass 0.1
of our $8^4$ simulation we did not find any zero modes after equilibration. 

In order to understand this feature, which seems to contradict other findings
\cite{CuKrFr04,DeSc04} we also ran simulations with the Wilson Dirac operator
on a $4^4$  lattice. Here we did see would-be zero modes (i.e. reasonably small
real eigenvalues) quite frequently. This may be due to  the fact that the
fluctuations of the Wilson Dirac operator seem to be much larger than the
chirally improved operator. In fact with our algorithm we were not able to
simulate the Wilson Dirac operator on the larger $8^4$ lattice.  The
fluctuations of the  fermionic determinant were far too large for any
reasonable trajectory length.

As a further check we also replaced the stochastic estimator for the
determinant by the exact evaluation for the small lattice runs. The tunneling
behavior did not change.  Even when we started with a topologically non-trivial
quenched configuration (i.e., zero modes of \DCI for the quenched gauge
configuration) the configurations quickly tunneled to the zero-topological
charge sector while performing the HMC updates. We conclude that there is no
obvious tunneling barrier in our algorithm but that the trivial zero mode
sector is natural for our choice of couplings.

Recent  results on the  dynamical  overlap fermions do report seeing zero modes
\cite{CuKrFr04,DeSc04}.  However, the plaquette values quoted in
\cite{CuKrFr04} are quite different from ours; indications are that those runs
are effectively at  smaller $\beta$ than ours. That coupling leads to a larger
physical volume (and lower temperature) and it is not obvious that the results
can be compared.   The results of \cite{DeSc04} extend to similar parameter
values as ours and tunneling to sectors with one zero mode were observed. We
work at a slightly larger gauge coupling, and, as argued below, we may be
deeper in the deconfined phase. More statistics and fine tuning of the
parameters will be necessary to resolve this discrepancy.

\subsection{Propagators}

\begin{figure}[tp] 
\begin{center}
\includegraphics*[width=6.5cm]{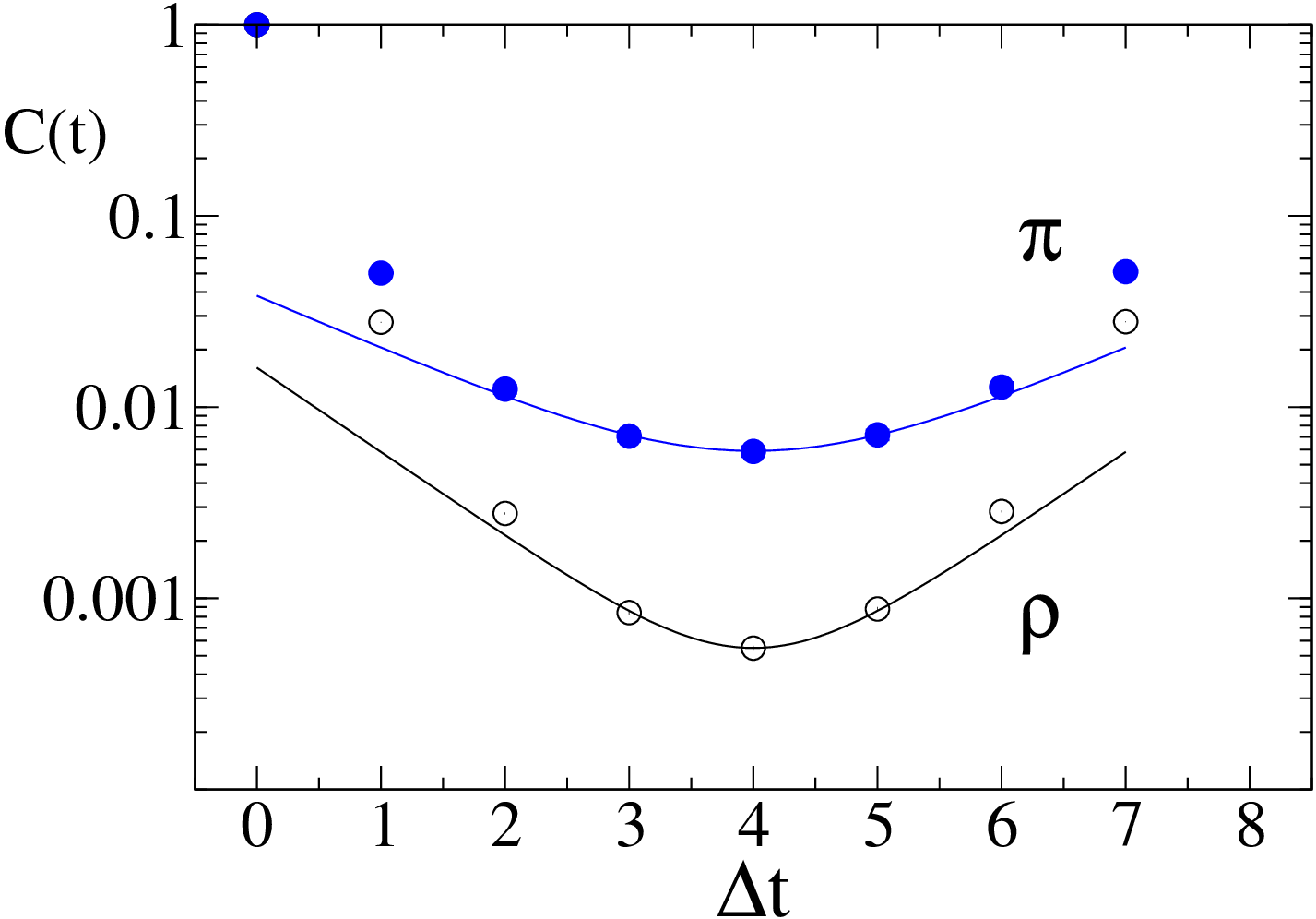}
\includegraphics*[width=6.5cm]{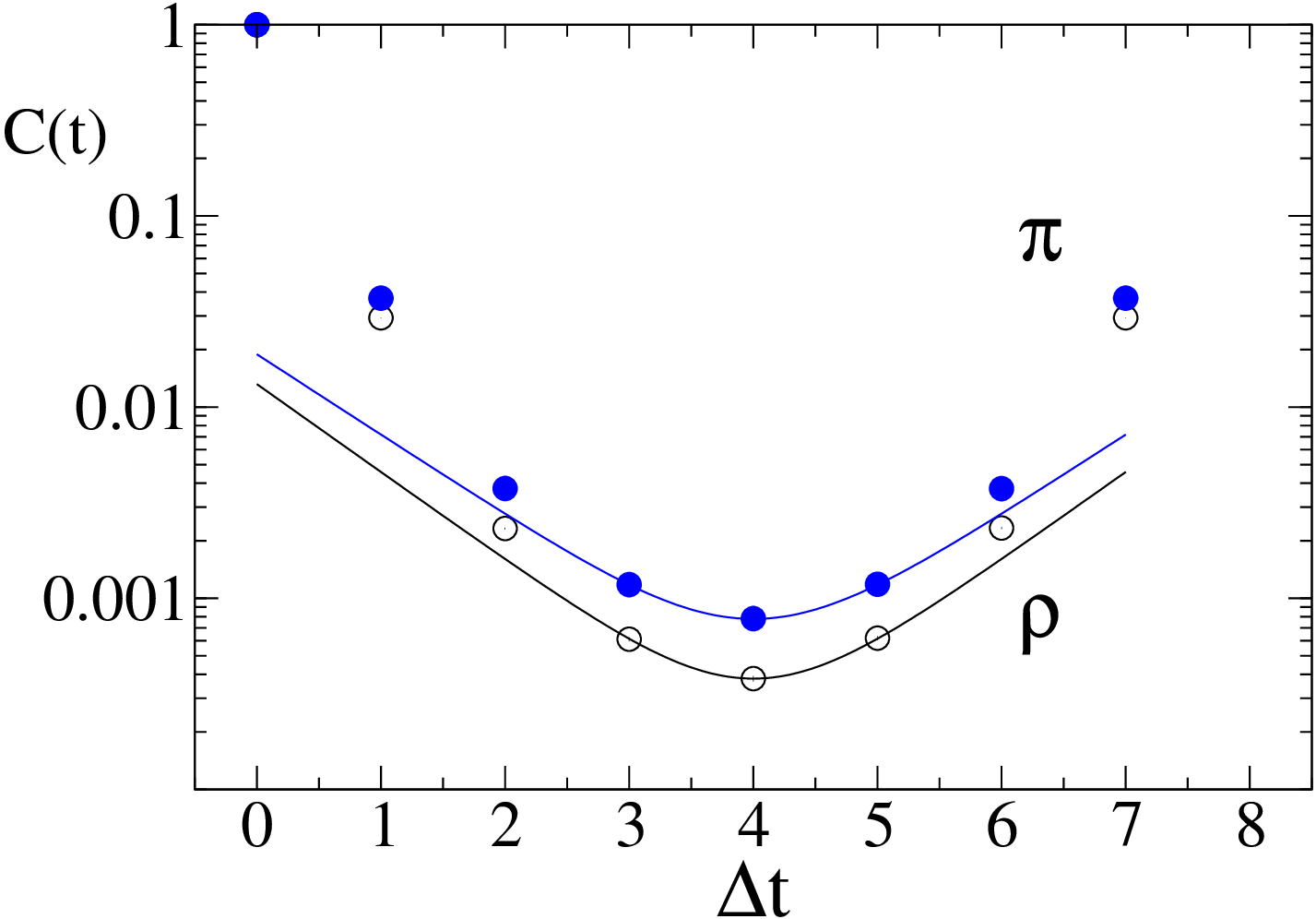}
\caption{Normalized pion (full circles) and rho (open circles) correlators on a $8^4$ lattice. 
We compare the quenched
results (left-hand plot)  with the dynamical ones (at $m_\srm{sea}=0.1$, right-hand plot),
both at $\beta_1=7.6$ and valence-quark mass $m_\srm{val}=0.1$. The curves represent the $\cosh$-fits
to the three central points.}\label{fig:correlators}
\end{center}
\end{figure}

One of the primary goals of lattice simulations is to reproduce the known hadron mass
spectrum. Although the lattice size is definitely too small to identify asymptotic states,
we still have computed
pion and rho meson propagators with our  dynamical chirally improved configurations for
a crude, preliminary check. The propagators were computed using point sources and
sinks. Our expressions for the pion and the rho correlators are given by
\bea
C_\pi(0,t) &=& \sum_{\vec x} 
\trace\left(\gamma_5\,\DCI^{-1}(\vec x,t:0,0)\,\gamma_5\,\DCI^{-1}(0,0:\vec x,t)\right)\;, \\
C_\rho(0,t) &=& \sum_{\vec x, i=1,2,3} 
\trace\left(\gamma_i\,\DCI^{-1}(\vec x,t:0,0)\,\gamma_i\,\DCI^{-1}(0,0:\vec x,t)\right)\;.
\eea
The correlation functions are shown in Fig. \ref{fig:correlators} for the dynamical 
($m_\srm{sea}=m_\srm{val}=0.1$) as well as for the quenched ($m_\srm{val}=0.1$)
case. The error bars are naively determined without auto-correlation analysis.

\begin{table}[t]
\caption{Fitted mass values in lattice units for lattice size $8^4$, $\beta_1=7.6$ and
valence-quark mass
$m_\srm{val}=0.1$.}\label{tab:tab1}
\begin{center}
\begin{tabular}{c|c|rr|rr}
          &          & \multicolumn{2}{|c}{Pion} & \multicolumn{2}{|c}{Rho} \\
\hline
simulation& range of $\Delta\,t$ & $m_\pi$ & $\chi^2/d.o.f$ & $m_\rho$ & $\chi^2/d.o.f$ \\
\hline
dynamical & 2$-$6   & 1.18 (3)   & 10.54 & 1.33 (4)  & 10.62 \\
          & 3$-$5   & 0.97 (1)   & 0.02  & 1.06 (2)  & 0.14 \\
\hline
quenched  & 2$-$6   & 0.71 (1)   & 0.13  & 1.21 (4)  & 8.16 \\
          & 3$-$5   & 0.64 (2)   & 0.03  & 1.02 (6)  & 1.62  \\
\hline
\end{tabular}
\end{center}
\end{table}

Some conclusions can be drawn comparing the propagators of the dynamical with those
of the quenched case. Table \ref{tab:tab1} shows the results for the ``meson masses''
$m$ of  our $\cosh(m\,(t-n_T/2))$-fits in the symmetry region. Let us denote the
dimensionless masses by $m^q$ (for quenched) and $m^d$ (for dynamical) and use the
values from the smaller fit range (with better $\chi^2/d.o.f$). We find
$m_\pi^d/m_\pi^q \approx 1.52$ and $m_\rho^d/m_\rho^q \approx 1.04$. There can be two
reasons for this. Either the lattice spacing has increased in the dynamical case or
it has decreased so much that we are at much smaller physical time extent and
therefore  cannot observe the asymptotic decay of the correlators. Our spectrum, as
discussed in the previous section, suggests that the second explanation is more
plausible. Thus we cannot use these values to derive the lattice spacing.

On the other hand, comparing the ratio of the fitted ``masses'' $m_\pi/m_\rho$ for
the dynamical and the quenched cases (0.92 vs. 0.63) we see that the dynamical ratio
is much higher. In the quenched case such a behavior  is observed for increased
valence-quark masses. Since the dimensionless valence-quark mass had the same value
0.1  in both cases we can use the meson mass ratio to try to obtain an effective
quenched $\beta_1$ (or lattice spacing) for our dynamical simulation. Results from
the BGR-collaboration's \cite{GaGoHa03a} quenched studies at values up to
$\beta_1=8.7$ (lattice spacing 0.078~fm) leads us to crudely estimate the gauge
coupling to lie beyond 8.7, corresponding to a lattice spacing $a \lesssim 0.08$~fm.

\subsection{Polyakov loop}

\begin{figure}[t]
\begin{center}
\includegraphics*[width=8cm]{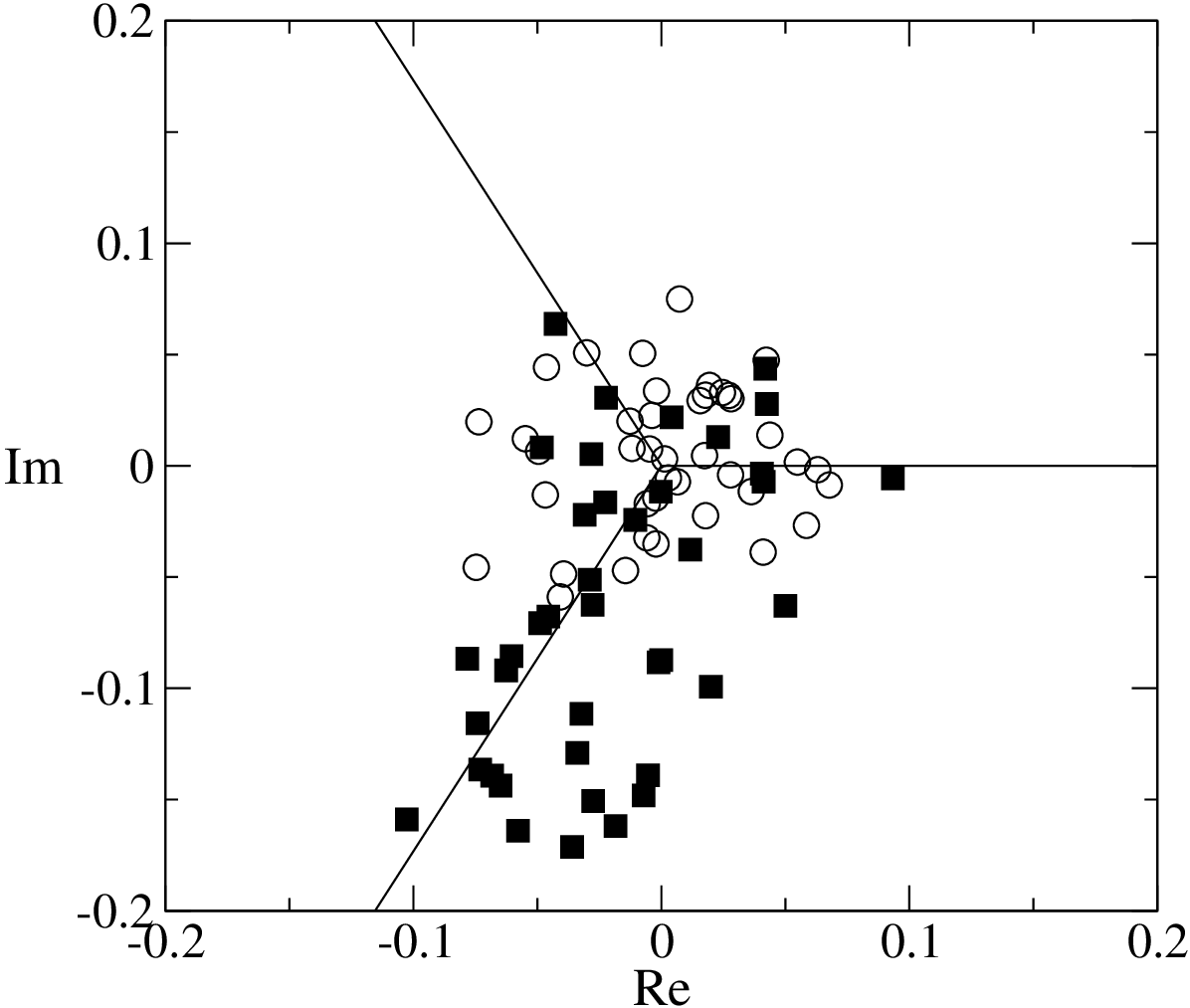}
\caption{Scatter plot of the Polyakov loop for the quenched (open symbols) and the
dynamical  (full symbols, $m_\srm{sea}=0.1$) situation for lattice size  $8^4$ and
$\beta_1=7.6$.
\label{fig:polyakov}}
\end{center}
\end{figure}

For dynamical quarks the Polyakov-loop is not an order parameter and for intermediate
values of the dynamical quark mass there may be no phase transition at all but just a
crossover region, analytically connecting the confinement region with the plasma
phase. Nevertheless, the Polyakov-loop is at least  an indicator for the location of
that crossover region. In Fig. \ref{fig:polyakov} we compare the values obtained for
the quenched and the dynamical case. We find that the center symmetry of the
confinement phase is broken for the dynamical situation.  This confirms our argument
that the effective lattice spacing has decreased considerably.

Since we are working on a symmetric lattice of size $8^4$ we are not really  entitled
to call $T=1/(8\,a)$ a temperature in a strict sense. However, assuming a transition
temperature range of $T_c\approx$  250~MeV  (quenched calculations give $T_c\approx$ 
270~MeV, dynamical simulations at smaller quark mass give  $T_c\approx$ 170-190~MeV
\cite{Ka00}) and still comparing it with $T$ we find that $a \leq 0.1$~fm for our
dynamical situation. This is significantly smaller than the quenched value
$0.19(1)$~fm.

This explains the missing zero modes. The dynamical system is too small to
accommodate  instantons and is already in the plasma-like regime.

\section{Conclusions}

We conclude that introducing dynamical fermions for chirally improved  ferm\-i\-ons
is possible with reasonable effort. HMC in the simplified version appears to work,
and dynamical fermions, although still with relatively large mass, make a difference
in the results as compared to the quenched case.

For the gauge coupling used we find evidence that the effective lattice  spacing
decreases considerably when switching on dynamical fermions. Although with our data
we cannot derive  values for the lattice spacing we have several observations
indicating the drastic decrease:
\begin{itemize}
\item The  Dirac operator spectrum  becomes smoother and closer to the GW-circle. 
Tunneling from non-trivial topological sectors to the trivial one is observed, but
the system then stays in the sector without zero modes.
\item The mass ratio of pion over rho  as derived approximately from the correlation
functions becomes larger. Since the dimensionless valence-quark mass is kept constant
this corresponds to a larger physical valence-quark mass but smaller lattice spacing,
similar to the observations in the quenched system.
\item The Polaykov loop shows breaking of the center symmetry.
\end{itemize}
All these effects are observed when increasing $\beta_1$ in a quenched simulations.
Curves of constant physics in the $(\beta_1, m_\srm{sea})$ plane bend towards smaller
gauge coupling for decreasing $m_\srm{sea}$. We estimate that the effective lattice
spacing has changed from 0.19~fm for the quenched simulation at  $\beta_1=7.6$ to a
value below 0.1~fm for $m_\srm{sea}=0.1$.

Obviously we should work on lattices with larger time extension for better analysis
of the propagators and  at smaller $\beta_1$ as a next step.  Also desirable is
an estimate of the growth of the computational effort with decreasing sea-quark mass.
In view of our results and within these caveats we can be optimistic to apply the
chiral improved fermion action to a realistic simulation of QCD.

{\bf Acknowledgment:}
We wish to thank Christof Gattringer for helpful comments; we are also grateful to
Nigel Cundy, Stefan Schaefer and  Peter Weisz for discussions.
Support by Fonds zur F\"orderung der Wissenschaftlichen Forschung in \"Osterreich,
projects M767-N08 (Lise-Meitner Fellowship) and P16310-N08 is gratefully acknowledged.

%%%%%%%%%%%%%%%%%%%%%%%%%%%%%%%%%%%%%%%%%%%%%%%%%%%%%%%%%%%%%%%%%%%%%%%%%%%
% BIBLIOGRAPHY
%%%%%%%%%%%%%%%%%%%%%%%%%%%%%%%%%%%%%%%%%%%%%%%%%%%%%%%%%%%%%%%%%%%%%%%%%%%

\end{document}